\documentclass[]{asa}

\usepackage{amsmath}
\usepackage{amssymb}
\usepackage{amsthm}
\usepackage{url}
\usepackage{xspace}
\usepackage{pdfpages}
\usepackage{color,soul}
\usepackage{setspace}

\newcommand{\R}{\textsf{R}\xspace}
\newcommand{\RM}{\textsf{R Markdown}\xspace}
\newcommand{\RStudio}{\textsf{RStudio}\xspace}
\newcommand{\knitr}{\textsf{knitr}\xspace}

\definecolor{Light}{gray}{.70}

\newcommand{\blind}[2]{#1} 

\title{R Markdown:\\Integrating A Reproducible Analysis Tool\\into Introductory Statistics}
\author{Ben Baumer\thanks{Department of Mathematics \& Statistics, Smith College, Northampton, MA 01063}
, Mine \c{C}etinkaya-Rundel\thanks{Department of Statistical Science, Duke University, Durham, NC 27708} 
, Andrew Bray\footnote[1]{Department of Mathematics \& Statistics, Smith College, Northampton, MA 01063} 
, Linda Loi\footnote[1]{Department of Mathematics \& Statistics, Smith College, Northampton, MA 01063} 
~and Nicholas J. Horton\thanks{Department of Mathematics, Amherst College, Amherst, MA 01002} }

\begin{document}

\begin{abstract}
Nolan and Temple Lang argue that ``the ability to express statistical computations is an essential skill." A key related capacity is the ability to conduct and present data analysis in a way that another person can understand and replicate. The copy-and-paste workflow  that is an artifact of antiquated user-interface design makes \emph{reproducibility} of statistical analysis more difficult, especially as data become increasingly complex and statistical methods become increasingly sophisticated.  \RM is a new technology that makes creating fully-reproducible statistical analysis simple and painless. It provides a solution suitable not only for cutting edge research, but also for use in an introductory statistics course. We present evidence that \RM can be used effectively in introductory statistics courses, and discuss its role in the rapidly-changing world of statistical computation. 
\end{abstract}


\section{Introduction}

Statistical analysis of data is both increasingly common and increasingly sophisticated. While the imperative to convey findings with clarity remains, the modern statistical analyst faces a variety of challenges that may make analyses more difficult to understand. First, as the field of statistics deepens, applications of statistics are increasingly complex. Second, collaboration among researchers is now the norm, rather than the exception. Third, much of that collaboration is conducted remotely, with written analyses, data files, and computing scripts shared via electronic means. Fourth, the underlying data being analyzed is larger and more complex, making it impossible to fully describe on paper, and thus necessitating transmission via an electronic file. Each of these complications makes it harder to completely follow someone else's work. Yet this task---in a word, \emph{reproducibility}---remains the lifeblood of scientific collaboration. 

In the past few years, the startling realization that many modern scientific findings \emph{cannot} be replicated has been highlighted in the popular press~\citep{Economist2013,johnson2014} as well as the scientific literature~\citep{ioannidis2013}.   Many factors have been identified, including publication bias, reporting bias, conflicts of interest, and insufficient statistical power.  This last factor can be remedied by encouraging the replication of studies and then conducting subsequent meta-analyses.  In order for a scientific study to be replicated, however, the method of statistical analysis must be entirely reproducible. Teaching reproducible analysis in an introductory statistics course not only makes students aware of these issues, but also paves the way toward making them valuable contributors to modern data analysis. These future contributions could be made as part of academic research or for a data-centric enterprise that needs to conduct daily analysis on new data.

The journal \emph{Nature} addressed these issues head-on in an editorial outlining the efforts that the journal would take to reduce their irreproducibility~\citep{editorial2013}.  A key provision is their pledge to ``examine statistics more closely and encourage authors to be transparent, for example by including their raw data."  In searching for the source of this irreproducibility, they note that ``mentoring of young scientists on matters of rigour and transparency is inconsistent at best" (p. 398).  A natural environment to provide this mentoring is the first time that most young scientists will encounter the formal principles of scientific inquiry: in introductory statistics.

The introductory statistics course has changed greatly in recent decades, with more focus on active learning, use of technology for conceptual understanding and analysis of data, along with ``naked, realistic and real'' data \citep{gaise}. In this paper, we discuss how \RM~\citep{markdown}, a simple, easy-to-learn, open source markup language, can be integrated into an introductory statistics course in an effort to achieve the GAISE guidelines, specifically, to enable students to develop the basic capacity to undertake modern data analysis and communicate their results. 

Our world is increasingly awash in data. What we plan to do with that information---how we plan to store it, how we will analyze it, and what exactly we hope to extract insight from it---are central unanswered questions facing data scientists, network scientists, statisticians, and computer scientists alike, both inside of academia and out. As statistics instructors, we face the difficult task of preparing our students to make their way in a sea of data that is foreign to many of us. Meanwhile, intrepid graduates hop on massive vessels like Google, Facebook, and Amazon each year, perhaps not knowing that the two-sample $t$-test they spent so much time studying in college is unlikely to be sufficient for their future work~\citep{cobb:2007}. But our aim here is not to discuss the \emph{topics} in introductory statistics courses, but rather, the \emph{workflow}. 

Hopefully the days in which students perform statistical analyses by hand or with a calculator are, if not over, bounded above by a function tending to zero. Thus, we take it as a given that our students' future work will be done on a computer. As such, providing students with the tools to ``think with data" and ``compute with data" is essential to their prosperity as a data analyst. At the same time, the ability to communicate one's findings to other people is imperative. At the end of \emph{every} data analysis task, there is a person who wants to understand the analyst's findings. That person may be a boss, a journalist, a doctor, or a policy-maker, who more often than not will have a weaker technical background than the analyst. This is the way it \emph{should} be, since otherwise, the person might as well perform the analysis themselves. But the end result is that the analyst's value is ultimately tied up in how understandable she can make her work. 

Because both computation and presentation are essential, a typical workflow is comprised of at least two major components: a statistical software package for performing the data analysis; and a layout package for presenting the results. For the former, we have had very positive experiences using \R~\citep{r-cran}, even in an introductory course, but other options (e.g. \textsf{SAS} or \textsf{Stata}) may be feasible.\footnote{Here we must distinguish between command-driven software packages (each of the aforementioned) and menu-driven software packages (e.g. StatCrunch or Microsoft Excel). It is increasingly the case that the complexity of data analysis tasks require the additional functionality and programmability of command-driven applications. The iteration required of students doing inquiry-based projects often breaks down in a menu-driven workflow.} For presentation, written reports tend to be composed in a word-processing application (e.g. Microsoft Word, LibreOffice Writer, or Google Docs) while oral presentations tend to use slides prepared in a presentation application (e.g. Apple Keynote, Microsoft PowerPoint, \LaTeX~with \texttt{beamer} or Prezi). A pairing of a statistical package and a layout package constitute a \emph{workflow}. That is, the analyst's work will typically begin in their statistical package of choice, wherein the data analysis will be performed. Once completed, translated summaries of work, be they tables, charts, images, or other output, need to be integrated into the layout application. Once in this environment, additional material can be layered onto the statistical results, so that a (usually less technical) human being can understand the findings.

This workflow is ubiquitous, and in most undergraduate courses where students are expected to compute with data, student homework assignments are produced in this manner. That is, statistical computations are performed in a statistics package, say \R, and then a written summary is produced in, say, Word. Tables and plots are simply copied-and-pasted from \R to Word. 

So if this workflow is so common, what is wrong with it? In truth, there are several important undesirable aspects. First, it is not reproducible. Since the commands used to generate the statistical output are not present in the final presentation, then either: a) the reader must assume that the student has calculated exactly what they say they have calculated, since there is no way of verifying the computation; or b) the grader must rely on the student to also copy-and-paste the commands used to generate the analysis. In either case, it will frequently be the case that the grader will be unable to completely follow the student's work. Moreover, the issue of reproducibility is relevant not only for a second-party (i.e. a grader), but also for the student. Being able to retrace steps while studying for a final, for example, is a desirable outcome. More concretely, the student may be reminded years later of the analysis, and seek to reapply the same methods in a different setting. Having the commands separated from the results inhibits this process. 

Second, the separation of computation from analysis is not logical. The commands in an \R script proceed chronologically, such that the analyst will most likely run the entire script all at once. A written report will be read in the same order, and there is no reason why the commands and analysis should not be interwoven. Rather, the impetus to separate the commands from the analysis is that the statistical package is not good at presentation, and the word-processing application is not good at computing. But this is not the ideal setup for the data analyst---it is simply an artifact of software design. \RM helps to bridge this gap in the data analysis workflow. 

Third, the separation of computing from presentation is not necessarily honest. At \blind{Smith}{XX} College, a strict honor code---to which all students are bound---discourages cheating. But it is all too easy for a student copying-and-pasting output from one program to another to fudge a few numbers. Again, the divorce of the computation from the presentation enables the student to edit the content along the way. The possibility of getting ``lost in translation" is disastrous for the data analyst. More subtly and less perniciously, the copy-and-paste paradigm enables, and in many cases even encourages, selective reporting. That is, the tabular output from \R is admittedly not of presentation quality. Thus the student may be tempted or even encouraged to prettify tabular output before submitting. But while one is fiddling with margins and headers, it is all too tempting to remove rows or columns that do not suit the student's purpose. Since the commands used to generate the table are not present, the reader is none the wiser.

Lastly, the copy-and-paste paradigm is error prone. When jumping between multiple windows ($\R$ and a word processor), students, often working on laptops with small cluttered screens, inadvertently copy-and-paste partial output or forget to update the output or plots included in the written report as they revise their analysis. This not only complicates grading, but it also results in increased frustration levels in students who devote time to improving their analysis but lose points for turning in a report that does not contain the desired results.

\section{Related Work}

While the notion that scientific results should be reproducible is fundamental, the recent interest in reproducible statistical analysis is a modern outgrowth fueled by developments in computing and networking. In particular, computational statistical methods have become more popular as computational power has become cheaper. Similarly, the Internet has eliminated many barriers to information dissemination. Thus, it is now possible to transmit the entirety of a statistical research project to nearly anyone in the world in almost no time, and at almost no cost. With the inclusion of both data and code the possibility exists that another person can entirely duplicate an analyst's findings with little effort. 

Knuth was an early advocate of \emph{literate programming}, which emphasized the use of detailed comments embedded in code to explain exactly what the code was doing~\citep{knuth1984literate}. The goal was to tie explanations to instructions so that work could be recreated, better understood, and verified. This idea was a predecessor to the notion of \emph{reproducible research}. According to~\cite{xie2014knitr}, the use of the term reproducible research first appeared in~\cite{claerbout1994hypertext}. Buckheit and Donoho were early disciples of Claerbout's ideas, incorporating them into their work with Matlab libraries~\citep{buckheit1995wavelab}. They proposed that in a scientific publication that relies on computation, the scholarship is not merely the presentation of the figures, etc. that further the author's case. Rather, ``the actual scholarship is the complete software development environment and complete set of instructions which generated the figures''~\citep{buckheit1995wavelab}. From here, it is clear that the burden of reproducibility rests on the original author, and that publication of  computer code is considered a necessary but not sufficient condition for achieving reproducibility.

Particular advocacy of reproducibility has come from the community surrounding \R. Sweave \citep{leisch2002sweave} provided a method for integrating executed \R code into \LaTeX~documents. The \knitr package by~\cite{xie2014knitr} provides equivalent functionality, but also partners with \RM to bring reproducibility and dynamic document generation to those who are not familiar with \LaTeX~\citep{gandrud2013reproducible}. In many ways \knitr can be seen as the realization of the vision for reproducible statistical analysis described by~\citet{gentleman2004statistical}. 

\nocite{hall2008introductory,fomel2009guest}

The emphasis on reproducibility can be seen as a necessary but not sufficient part of ensuring that students have capacity to ``think with data." Along these lines, recent efforts in statistics education have advocated for an increased use of computing in the statistics curriculum, both at the undergraduate and graduate levels~\citep{nolan2010computing}. Yet while they argue strongly for the need for students to learn programming (and presumably, literate programming) they provide no mechanism for allowing students to express their statistical computations. \RM provides exactly such a mechanism, and fits squarely into the statistical computing workflow. 
\section{R Markdown}

\RM is an easy-to-use system that enables students to combine statistical computing in an environment of their choosing and written analysis in \emph{one document}. At a high-level, it renders a well-annotated \R script into a self-contained HTML file, replete with graphics, commands, and stylized text. 

Like \LaTeX~or HTML, \RM relies on a \emph{source} file and \emph{output} file paradigm. Text, with simple rules for creating styles, is typed into an \RM source file, which has the \texttt{.Rmd} extension. \R commands are typed directly into this file, set off in ``chunks". The \knitr rendering engine then parses the \texttt{.Rmd} file. It first executes each of the \R commands in the chunks and processes the output from those commands. This generates an intermediate Markdown file (with a \texttt{.md} extension) which is of no immediate interest. Next, it renders this Markdown file into a single HTML file with 
embedded graphics. For those familiar with \LaTeX, Sweave, or PHP, it is very similar to the way that each of these process one source file into another output file. A comparison of the workflow in rendering applications is given in Table \ref{tab:render_app}.

\begin{table}[h]
	\centering
\begin{tabular}{|c|c|c|p{2cm}|p{2cm}|}
	\hline
	Source Language &  Source file format & Rendering Engine & Intermediate file format & Output file format \\
	\hline
	\LaTeX 	& .tex & pdflatex & .log, .aux & .pdf, .ps \\ 
	Sweave 	& .Rnw & Sweave, knitr & .tex	& .pdf \\
	PHP			& .php & PHP & & .html \\ 
	\RM			& .Rmd & knitr & .md & .html \\
	\hline
\end{tabular}
	\caption{Comparison of similar rendering applications}
\label{tab:render_app}
\end{table}

The primary benefit of \RM is that it restores the logical connection between the statistical computing and the statistical analysis that was broken by the copy-and-paste paradigm. Each chunk of \R code is rendered into two parts: first, a box that contains the syntax-highlighted, tidied \R code; followed by the output from those commands. In this manner, it is perfectly clear exactly what command has been run, and there is no way to fudge or edit the output from those commands.\footnote{OK, a hacker-student could edit the HTML file manually, but good luck trying to edit the figures, which are rendered as bytecode to allow them to be saved as embedded images! That is, instead of the typical configuration wherein images on a web page are stored in separate files, \RM converts all images to an equivalent HTML string of machine-readable bytecode. This allows each rendered \RM document to include images without requiring external files.} Additional content in the form of text, lists, headers, tables, external images, and web links, etc. can surround the \R chunks in a standard way.

One of the major advantages of \RM over existing technologies, such as Sweave \citep{leisch2002sweave}, is that the Markdown syntax is very simple. For example, to make a word show up in boldface, it is surrounded with asterisks. Compare this to HTML, in which you'd have to put ``<b>'' before the word and ``</b>'' after it. Or consider \LaTeX, in which you would have to encase the word: \verb+\textbf{word}+. A side-by-side comparison of the alternatives are shown in Table \ref{tab:word}.

\begin{table}[h]
	\centering
	\begin{tabular}{|c|c|c|}
		\hline
		HTML & \LaTeX & \RM \\
		\hline
		\texttt{<b>word</b>} &  \verb+\textbf{word}+  & \texttt{*word*} \\
		\hline
	\end{tabular}
	\caption{Comparison of syntax for typesetting ``word" in bold face. The syntax employed by \RM does not require learning a separate set of complex rules, as does HTML or \LaTeX.}
	\label{tab:word}
\end{table}

\begin{figure}[ht]
\centering
\includegraphics[width=\textwidth]{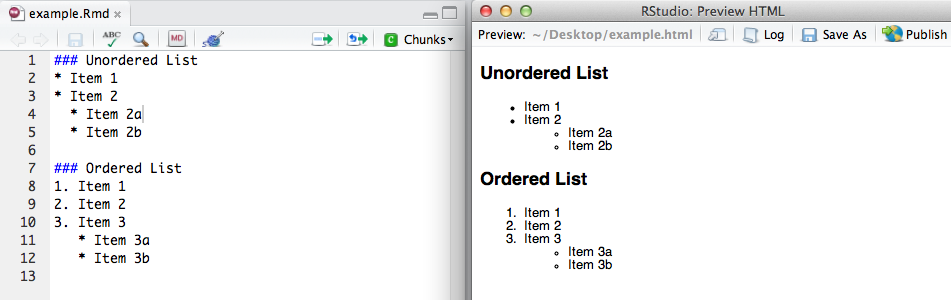}
\caption{Bulleted list in \RM, input (left) and output (right).}
\label{fig:list_example}
\end{figure}

To make a bulleted-list in \RM, a series of lines are prefaced with an asterisk in exactly the manner as in a plain-text email (See Figure~\ref{fig:list_example}). Thus, students can learn to use \RM without the burden of learning a wholly new technology, such as \LaTeX~or HTML. The \RM syntax is so simple that the majority of the \RM syntax is presented on a short web page~\citep{r-markdown}.

\begin{figure}[]
\centering
\includegraphics[width=\textwidth]{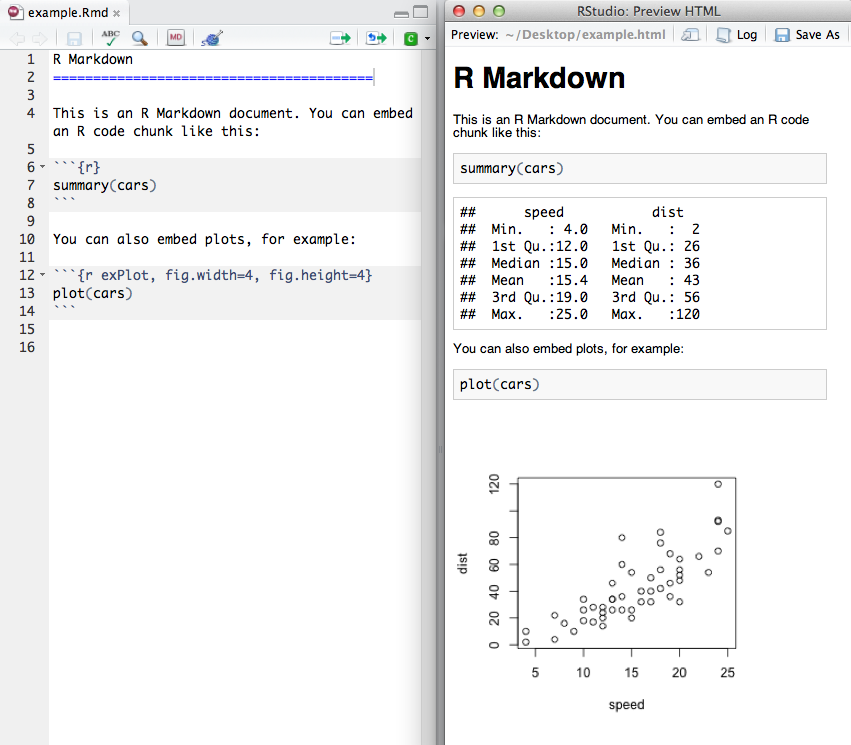}
\caption{An example of an \RM chunk (left) and its rendered output (right).}
\label{fig:chunk}
\end{figure}

\R commands and output are distinguished from plain text with the use of \emph{chunks}. Chunks begin with a series of three backticks, and conclude with three more. Figure \ref{fig:chunk} illustrates a simple chunk of \R code and its rendered output. Note that the chunk in Figure \ref{fig:chunk} is named (\texttt{exPlot}), and sets two options to non-default values (\texttt{fig.width, fig.height}). 

We should note that 
the \knitr rendering engine is not specific to \R or \RStudio, a popular open source integrated development environment for \R. The following \R commands are equivalent to clicking on the ``Knit HTML" button in \RStudio (note the intermediate generation of a Markdown file):
\begin{verbatim}
library(markdown)
knit("filename.Rmd")   # creates filename.md
markdownToHTML("filename.md", "filename.html")
browseURL("filename.html")
\end{verbatim}
\RStudio is available as either a client application or a server (cloud-based) version.  The latter setup, implemented at our institutions, allows students to access and run \RM and \RStudio through a browser, and minimizes startup time.


Moreover, while in this paper we focus on the use of $\RM$ in the introductory statistics class, we should also note that, just like $\R$, $\RM$ also extends beyond the introductory classroom. Students who are introduced to the concept of reproducibility at this level carry the skills they acquire with them throughout their undergraduate career (and beyond). At the point where the simple formatting of $\RM$ becomes limiting to producing high quality customizable reports, students who are familiar with \LaTeX~can easily transition to Sweave/\knitr. In fact, at \blind{Duke}{XX} University, students taking the Statistical Consulting course (STA \blind{470}{XX3}) as one of the last courses in the major curriculum use Sweave/\knitr to complete their data analysis assignments, as do students in Mathematical Statistics
at \blind{Smith}{XX} College.

\section{Using R Markdown in Introductory Statistics}

\blind{\subsection{Duke University}}{\subsection{XX University}}

At \blind{Duke}{XX} University in \blind{Durham, North Carolina}{City, ST}, 272 statistics students have used \RM during the 2012-2013 academic year (221 enrolled in STA \blind{101}{XX1} during the Fall and Spring semesters, and 51 enrolled in STA \blind{102}{XX2} during Spring). Both of these are non-calculus based introductory statistics courses usually taken by first and second year students majoring in either the social sciences or the life sciences, respectively. Only a very small number of the students each semester have any meaningful computational background. Both courses have lecture and lab components, and students used \RM to complete their lab assignments as well as data analysis project(s). In STA \blind{101}{XX1}, students complete a simpler project on statistical inference evaluating univariate distributions or bivariate relationships (completed individually) and a more advanced project on multiple regression (completed in teams). In STA \blind{102}{XX2}, the students complete an open-ended data analysis project using the appropriate methods covered in the course (competed individually).

The STA \blind{101}{XX1} course employs the flipped classroom model as well as team-based learning. Students are assigned to teams by the instructor at the beginning of the semester based on their performance on the ARTIST CAOS, Comprehensive Assessment of Outcomes in a First Statistics course, \citep{delm:2007}, pre-test and their responses to a survey on their statistics, mathematics, and computer science background as well as their interests and reasons for taking the course. The teams are created to be heterogenous with respect to statistics experience and homogenous with respect to student interests. Students work in teams in many components of the course, including the weekly \R labs. The final product of the weekly labs is a team lab report, created using \RM. The labs are designed such that students complete the majority of the exercises during the lab sessions led by the teaching assistants. However, finalizing the analysis and the write-up requires spending time outside of class. Reports produced in \RM facilitate easy and organized sharing of the code and the write-up among the team members. Prior to integrating \RM into the course curriculum students struggled with sharing their work among team members and version control. Using \RM for the weekly labs allows students to work collaboratively on data analysis throughout the semester, and they reap the benefits of having developed a workflow that has reproducibility at its heart when working on their larger scale individual and team projects. 

In addition, reports produced using \RM present the code and the output in one place (as input and output) making it easier for students to learn \R and locate the cause of an error. Likewise, uniformity of the output and the enforced structure of the reports significantly aid the instructors in debugging issues as they arise as well as simplifying the task of grading (see Appendix \ref{sec:sample} for a sample lab assignment and student solution).

In previous versions of the course, prior to adopting \RM, labs and projects still required analyses performed in \R. As the students were learning \R concurrently with new statistical concepts, they would often struggle to organize their analyses. They took a trial-and-error approach to coding, and made ad-hoc changes as they went through the analysis. However, despite trying to instill best practices, most students never really developed a habit of separately saving their code. This often resulted in cluttered workspaces and \R consoles, difficult-to-diagnose errors due to overwriting data, and hence student frustration. We believe that the root of the problem was that the desired final product (the lab report, the project write-up, etc.) was just a presentation of results (typed up in a Word processor like Microsoft Word or Google Docs) that did not include the underlying code. On the other hand, comments from students enrolled in recent versions of the course, after adopting \RM, suggest that they appreciate the ease of organization of their code:

\begin{itemize}

\item ``I think the labs have been great. Using \RM has been so great because we do not spend as much time solving the format/design of the paper and instead focus on actual problem solving. \R is super easy to use and useful. "

\item ``The labs have been enjoyable, and \RM makes the process very easy."

\item ``Labs can sometimes be troublesome and confusing, however, the TAs are very helpful. The \RM used to generate lab reports/proposals are very helpful for organizing our information."

\end{itemize}

Students also commented on the usefulness of templates provided with the labs (see Appendix~\ref{sec:sample}). Another notable point was a general sense of excitement and interest around the labs.

\begin{itemize}

\item ``The labs are fun.  There is something satisfying about hitting `knit' and having the text turn into figures and tables."

\item ``I like it a lot actually. It has sparked an interest in coding for me."

\end{itemize}

\blind{\subsection{Smith College}}{\subsection{XX College}}

At \blind{Smith}{XX} College in \blind{Northampton, Massachusetts}{City, ST}, 145 statistics students used \RM during the 2012-2013 academic year. In the fall semester, 42 students completed MTH \blind{245}{XX1}, an advanced first course in statistics for students with a calculus background. The course is worth five credits and has both lecture and lab components. These students completed most of their lab assignments in \RM. Furthermore, after conducting a statistical investigation involving multiple regression as part of their final project, they submitted a ``technical appendix" composed in \RM. A total of 33 other students took a second course in statistics, MTH \blind{247}{XX2}, which focused on regression analysis. These students completed all of their homework assignments in \RM and for their final project, submitted both a technical appendix written in \RM and a write-up composed in a word-processing application. 

Anecdotal success with this pilot program at \blind{Smith}{XX} led to the integration of \RM into three sections of the spring semester introductory statistics course. \blind{241}{XX3} is the four credit equivalent of MTH \blind{245}{XX1}, which similarly requires calculus but does not have a lab component. As in MTH \blind{245}{XX1}, 70 students completed almost all homework assignments in \RM, as well as a technical appendix for their final project. These students were given surveys at the beginning and end of the semester in order to gauge their attitudes toward \R and \RM. (This project was approved by the \blind{Smith}{XX} College Institutional Review Board.) The results, which we present in detail below, suggest that:

\begin{enumerate}
	\item Students grew to appreciate \RM's ability to streamline their homework workflow. In particular, students did not prefer to copy-and-paste their work from \R into Microsoft Word. 
	\item While students experienced frustration with both \R and \RM, this frustration waned over the course of the semester.
	\item There was little to no correlation between a student's attitude towards \RM and that student's performance in the course.
	\item Lack of prior exposure to markup languages similar to \RM was not an impediment to learning \RM.
\end{enumerate}

From the point of view of the instructor, while there is some overhead and growing pain required alongside the introduction of \RM, these hurdles will be overcome, and the benefits are well worth it. Specifically, the lesson of reproducibility is emphasized throughout the semester, homework is easier to grade, and students receive more comprehensive and specific feedback on their statistical computing than they would using the typical copy-and-paste paradigm. 

\subsubsection{Survey data}

\begin{figure}
	\centering
	\includegraphics[width=\textwidth]{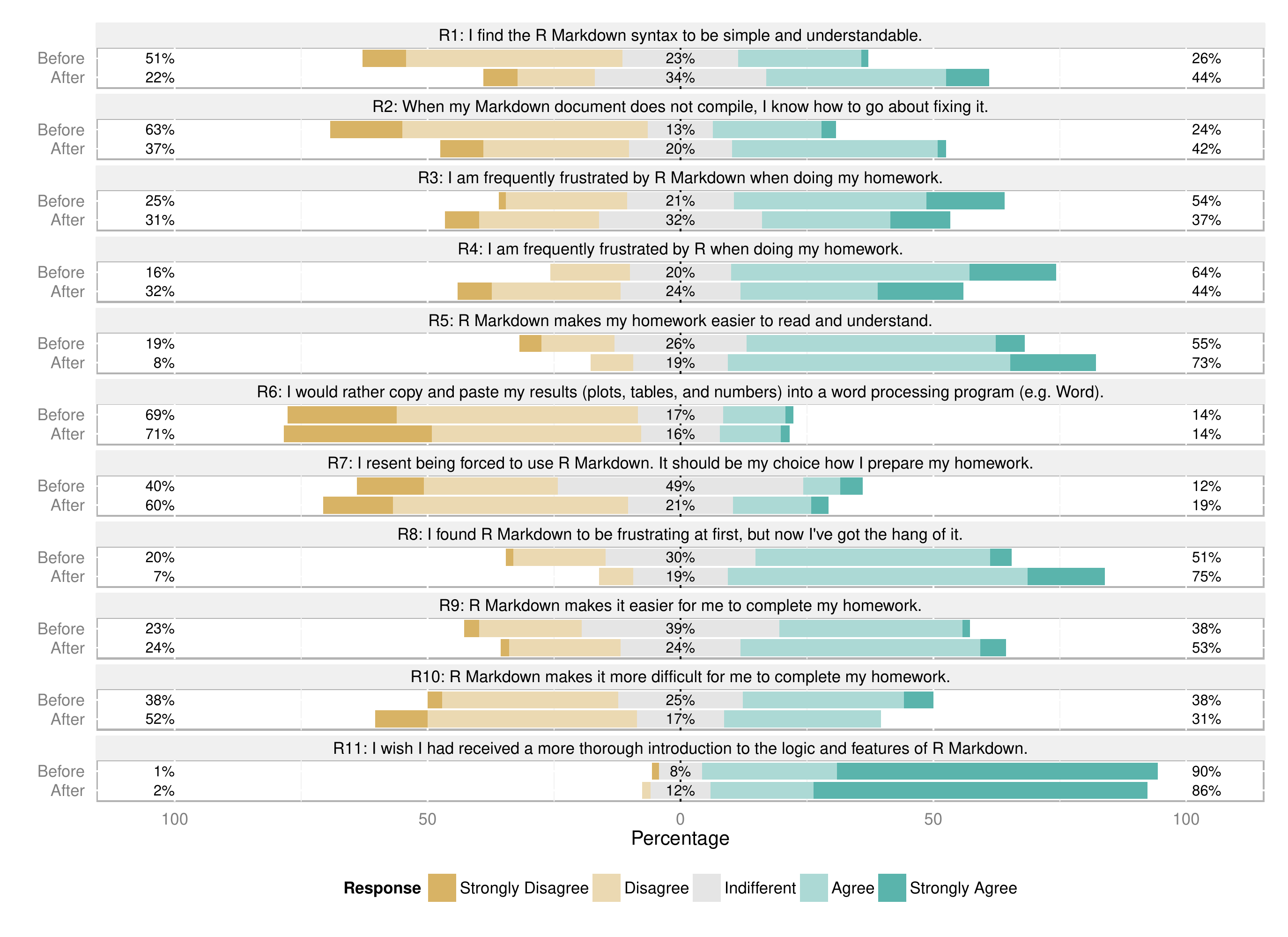}
	\caption[Likert Results]{Results from Likert scale \RM survey administered at \blind{Smith}{XX} College, also summarized in Table \ref{tab:responses}. Responses from students who circled more than one answer were rounded to the more extreme value.}
	\label{fig:likert}
\end{figure}

Of the aforementioned 70 students, 56 completed the Likert-scale survey shown in Appendix~\ref{sec:survey} at both the beginning of the semester (after some initial exposure to \R and \RM), and at the end of the semester. A summary of their responses to questions is shown in Table \ref{tab:responses} and Figure \ref{fig:likert} \citep{likert}. The responses in Table \ref{tab:responses} are scored on a scale from $-2$ to 2, where $-2$ represents strong disagreement with a statement that is favorable to \R or \RM, and 2 represent strong agreement with that same statement. Note that only about half of the statements on the survey were worded favorably towards \RM, so for questions 3, 4, 6, 7, 10, and 11, $-2$ corresponds to ``strongly agree", whereas for questions 1,2,5,8, and 9, $-2$ corresponds to ``strongly disagree". Thus, in the interpretation of Table \ref{tab:responses}, larger positive numbers are always good (from the point-of-view of \RM).

\begin{table}[ht]
\centering
\begin{tabular}{c|c|cc|cc|cc}
 & & \multicolumn{2}{|c|}{Before} & \multicolumn{2}{|c|}{After} & \multicolumn{2}{|c}{Change} \\
  \hline
Question & Idea & N & Mean (SD) & N & Mean (SD) & N & Mean (SD) \\ 
  \hline
B1 & prior \R &  56 & 1.30 (0.60) &  56 & 1.34 (0.58) &  56 & 0.04 (0.50) \\ 
  B2.CSS & prior CSS &  56 & 0.14 (0.35) &  56 & 0.12 (0.33) &  56 & -0.02 (0.23) \\ 
  B2.HTML & prior HTML & 56 & 0.46 (0.50) &  56 & 0.48 (0.50) &  56 & 0.02 (0.45) \\ 
  B2.LaTeX & prior \LaTeX & 56 & 0.07 (0.26) &  56 & 0.07 (0.26) &  56 & 0.00 (0.19) \\ 
  B2.Wiki & prior Wiki & 56 & 0.14 (0.40) &  56 & 0.12 (0.33) &  56 & -0.02 (0.40) \\ 
  B2.XML & prior XML & 56 & 0.00 (0.19) &  56 & 0.04 (0.19) &  56 & 0.04 (0.19) \\ 
  R1 & simplicity & 55 & -0.30 (0.93) &  56 & 0.24 (1.05) &  55 & 0.53 (1.12) \\ 
  R2 & compilation & 55 & -0.53 (1.07) &  56 & -0.04 (1.05) &  55 & 0.48 (1.42) \\ 
  R3 & RM frustration & 56 & -0.50 (1.04) &  56 & -0.10 (1.07) &  56 & 0.40 (1.19) \\ 
  R4 & \R frustration & 55 & -0.68 (0.90) &  56 & -0.21 (1.17) &  55 & 0.45 (1.26) \\ 
  R5 & readability & 55 & 0.35 (0.95) &  56 & 0.84 (0.80) &  55 & 0.51 (0.79) \\ 
  R6 & copy-and-paste & 51 & 0.73 (0.94) &  55 & 0.87 (1.06) &  50 & 0.10 (0.99) \\ 
  R7 & coercion & 53 & 0.35 (0.99) &  55 & 0.55 (1.02) &  52 & 0.24 (1.03) \\ 
  R8 & improvement & 56 & 0.22 (0.83) &  56 & 0.83 (0.75) &  56 & 0.61 (0.94) \\ 
  R9 & ease & 55 & 0.08 (0.89) &  56 & 0.33 (0.93) &  55 & 0.25 (1.03) \\ 
    R10 & difficulty &  55 & -0.05 (1.00) &  55 & 0.30 (1.00) &  55 & 0.35 (0.99) \\ 
  R11 & training & 56 & -1.46 (0.79) &  56 & -1.51 (0.79) &  56 & -0.04 (0.95) \\ 
   \hline
\end{tabular}
\caption[Responses to Questionnaire]{Summary of before and after responses to questionnaire. Responses were scored according to the scale: no opinion $=$N/A, strongly disagree $=-2$, disagree $=-1$, indifferent $=0$, agree $=1$, strongly agree $=2$. The responses to questions 3, 4, 6, 7, 10, and 11 have been flipped. Thus, higher scores are more favorable to \R and \RM, and lower scores are less favorable. Note that what is being shown in the third group of columns is the mean change in response, not the change in mean response.} 
\label{tab:responses}
\end{table}

%

Questions 5, 6, 7, 9, and 10 address \RM's role in the data analysis workflow. For all five questions, the students responses were favorable at the end of semester, and grew more favorable over the course of the semester. Most notably, while students were largely indifferent to \RM's ability to make their homework easier to read and understand at the beginning of the semester (mean initial response to R5 of 0.35), by the end of the semester the most students realized this benefit (mean final response to R5 of 0.84). The improvement of 0.51 was among the largest changes for any of the eleven questions. Note that this question forces the students to consider the perspective of someone reading their work---it does not solely address a question in the student's immediate self-interest.

Moreover, while the initial response to questions 9 and 10 were indistinguishable from zero, by the end of the semester there was mild agreement that \RM makes it easier for students to complete their homework. Thus, students acknowledged that \RM, in addition to being a benefit to their audience (as demonstrated by question 5), was of a mild benefit to them. 

Questions 6 and 7 address the possibility of alternative workflows. In question 7, students expressed a mild lack of resentment at being forced to use \RM. However,  residual resentment waned over the course of the semester. More interestingly, students were quite opposed to the typical workflow which would require them to copy-and-paste their results from \R into Word. Moreover, there was little change in these responses over the course of the semester. Thus, the results suggest that not only do students prefer \RM to Word after having used it all semester long, but that they never preferred to use Word in the first place. This should help to encourage those instructors who are most comfortable in Word to consider making a change. It does not appear that these students were beholden to word processing applications. 

Questions 3, 4, and 8 address the issue of frustration with \R and \RM. Here, it is expected that many students will express frustration with \R, which is an admittedly expert-friendly software package. The data suggests that while initial frustration with both \R and \RM was reasonably high, by the end of the semester it had largely dissipated. In particular, frustration with \RM was negligible by the end of the semester, and frustration with \R was considerably diminished. This notion was addressed more directly by question 8, which offered the largest change over the course of the semester (0.61). Here, most students agreed that they were frustrated by \RM at first, but had gotten the hang of it by the end of the semester. 

Questions 1, 2, and 11 address the students' experience working with \RM. On Question 11, students were almost unanimous is their desire to have received a more thorough introduction to the logic and features of \RM. Unlike the previous questions, this initial reaction was not moderated over the course of semester. While it is expected that many students will request additional help in working with new technologies, future versions of the course will include some kind of ``workshop" during the first month that eases the adoption of \R and \RM. On the other hand, questions 1 and 2 show evidence of student growth. While many students did not find \RM to be particularly simple and understandable upon initial exposure, by the end of the semester they mildly supported the claim that \RM was simple and understandable. Perhaps more importantly, students showed a marked improvement in their ability to debug \RM. At the beginning of the semester, many students did not feel as though they knew how to fix compilation errors in \RM, but by the end of the semester, they did not disagree (to a statistically significant extent) with the notion that they could debug their own \RM errors.

\subsubsection{Anticipated Problems}
\label{sec:cor}

Two fears that we had did not seem to be supported by the data. First, we feared that since the use of \RM was so thoroughly integrated into the course, and so vital for completing the homework (which constituted 20\% of the total grade for the course), that students who viewed \RM more favorably would be advantaged with respect to their overall grade in the course. Second, we feared that students who had stronger prior exposure to technologies similar to \RM would be have an easier time completing their assignments. More specifically, we feared that students who had not been exposed to technologies similar to \RM would suffer since they might have to spend more time on their homework. Neither of these fears were borne out in the data. 

To test these hypotheses, we examined the relationships between survey responses at the beginning and end of the semester, and two measures of performance in the course: the student's final course grade, and her score on the Comprehensive Assessment of Outcomes in a First Statistics Course (CAOS;~\cite{delm:2007}) post-test. None of the correlations between the scores on each of the 11 questions and the student's final course grade were statistically significant at the 5\% level~\footnote{We acknowledge that none of the measures of statistical significance reported were corrected for multiple comparisons. However, since the purpose of this analysis is to show that there is little statistical evidence of correlation between attitudes towards R Markdown and performance in the course, and a multiple comparisons correction would only \emph{weaken} any claims of statistical significance, we do not feel that this omission detracts from our findings.}. Only two (R5 and R6) were significant at the 10\% level, with both indicating weak positive association with final course grade (0.23 and 0.26, respectively). Correlations between the responses and CAOS scores revealed a similar lack of association (R6 showed a correlation of 0.27, but a 95\% confidence interval $[-0.004, 0.514]$ for the true value of this parameter still included 0). Moreover, assessment of the association between the two performance measures and the change in response over the course of the semester revealed no statistically significant associations. Thus, not only were the students' initial reactions to \RM not an important indicator of their performance in the course or their absorption of statistical material, but neither was their change in attitude towards \RM over the course of the semester. 

Similarly, prior exposure to \RM-like technologies did not appear to be associated with student performance. First of all, only one quarter (14 of the 56) of the students had ever heard of \R prior to taking the course, and only two had used it. Only four students had prior exposure to \LaTeX, and only nine reported having edited a Wiki. While eight students had seen Cascading Style Sheets (CSS), all eight had prior exposure to Hypertext Markup Language (HTML), along with 18 students who had used HTML but not CSS. Thus, prior exposure to HTML was the only prior technology to which students had reasonably varied backgrounds. While there was no association between prior exposure to HTML and score on the CAOS exam, there was a borderline statistically significant \emph{negative} correlation between prior exposure to HTML and final course grade ($p = 0.051$). Due to the marginal significance of this result, its counter-intuitive nature, and the multiple uncorrected tests we performed, we do not interpret it as being of interest. 

\subsubsection{Ancillary Outcomes}

Finally, the end-of-semester responses to questions 5 and 6 deserve a moment's reflection in their own right. For the most part, students agreed (0.84) that \RM made their homework easier to read and understand. [To what they were comparing it to, perhaps handwritten or pasted into a Word document, is left open.] Moreover, they would not rather (0.87) copy-and-paste their homework into Microsoft Word. We interpret the responses to question 5 as an affirmation of \RM's usefulness for students, and note that this perception grew over the course of the semester. The responses to question 6 confirm that working with \RM for a semester, and the occasional frustration that goes along with it, did \emph{not} make students yearn for a return to Word. While this attitude did not change much of the course of the semester, it reveals the perhaps surprising discovery that even students who have never heard of \R a few weeks before would not rather copy-and-paste their statistical results into Word as part of the homework preparation. We interpret these findings as further evidence that open-source tools are perfectly suitable for use in even introductory statistics courses at the undergraduate level.

\section{Discussion}

Having presented a motivation for using \RM in introductory statistics, described the technology, and reviewed our experience using it, we close with a discussion of some additional benefits, challenges, and limitations. 

\subsection{Challenges and common problems}

One of the benefits of using \RM in both the introductory and intermediate statistics courses is the development of knowledge within the institution. At \blind{Smith}{XX}, one of five statistics teaching assistants is available for two hours each night from Sunday to Thursday. All of these students are now familiar with \RM and capable of helping introductory students with common problems. In good faith, we present some of those issues below. 

\begin{itemize}
	\item Workspace confusion: Many errors result from a failure to understand that each \RM file, when compiled, runs in a fresh workspace that does not have access to any of the objects in the existing workspace active in \RStudio.	
	\begin{itemize}
		\item Failure to load packages: Students will often forget to load additional packages in their \RM scripts (e.g. \texttt{require(mosaic)}).
		\item Reading external data files: Students often forget to add the \texttt{read.csv()} in their \RM file after loading it into their workspace from typing it in the console.  	
	\end{itemize}
	\item Improper use of chunks: Students often forget to put their \R code into a valid chunk. A useful solution is to tell them to select ``Insert Chunk" from the green Chunks menu whenever they want to enter commands.
	\item Forgetting to close quotes or parentheses or chunks: Syntax highlighting in \RStudio mitigates this issue, but it still arises.
  \item Issues specific to \R as opposed to \RM: Invalid syntax for commands.
	\item Debugging: The compilation errors that occur when \RM is rendered are not always straightforward to interpret. Thus, students occasionally have a hard time identifying the particular command that is causing the problem. This can be mitigated by encouraging students to name their chunks, and to encourage them to pursue common process of elimination debugging techniques.  
	\item Package versioning: In some cases the packages on a students machine may become out-of-date or out-of-sync. Encouraging them to keep all of their packages up-to-date (especially \knitr) with the \texttt{update.packages()} command usually provides a solution. Alternatively, encouraging students to use a server version of \RStudio (administered by your institution) can be an effective solution.
	\item Formatting: While \RM is capable of implementing basic formatting operations, many more advanced features are not available. Some of the more useful and accessible options are:
	\begin{itemize}
			\item Gratuitous output: Without the \texttt{message=FALSE} option in an \RM chunk, unwanted messages are rendered in the output.
			\item Plot size: The size of a rendered plot can be changed by using the \texttt{fig.width} and \texttt{fig.height} chunk options.
			\item Chunk naming: Assigning a name to each \R chunk is helpful with debugging.
	\end{itemize}
	\item When all else fails: Restarting \RStudio can solve many problems. Any package can be safely removed and reinstalled. Occasionally doing this will solve less obvious problems. 
\end{itemize}

\subsection{Limitations}

While \RM is suitable for many purposes, it has a few limitations that may prove problematic. Specifically:

	\begin{itemize}
		\item While objects defined in previous chunks become part of the workspace and are thus available for later use, plots defined in previous chunks cannot be modified by later chunks. The most common work-around for this issue is to create a plot in a single chunk or assign the output of a plot to an object that can be printed in a subsequent chunk.
		\item There is no easy way to count words or pages in the rendered \RM output. This makes it difficult to check to see if a submitted homework assignment meets any such guidelines. 
	\end{itemize}

Use of the default formatting options in \RM can result in very long documents. If the rendered HTML file that a student wishes to submit is very long, then it can quickly become cumbersome and even expensive to print it out and submit a hard copy. On the other hand, if the document is to be submitted and evaluated electronically, the length of the document may be of no real concern. Thus, while use of even basic non-default formatting options can dramatically reduce the length of rendered \RM documents, there is a sense in which moving to electronic submission and grading will mesh well with \RM adoption. Indeed, if the grader knows HTML, it is even possible to give inline feedback on a student's submission. (This process has been implemented at two of our institutions.)

Given the interest in having students collaborate on projects at the introductory level~\citep{halvorsen2001motivating}, streamlining a collaborative workflow is worthwhile. \RM provides such a mechanism in part due to its inherent emphasis on reproducibility. Students working together are able to follow, and even extend, each other's work with minimal effort. Nevertheless, a fool-proof solution for having multiple students edit the same \RM file simultaneously does not yet exist.  The use of an \RStudio server, or a third-party file synchronization solution (e.g. Dropbox) can provide a functional workaround. Future versions of \RStudio may also include additional features designed to facilitate real-time collaboration projects. 

\subsection{Additional Thoughts}

Another component of reproducibility relates to the version of \R and its associated packages, which are often updated.  While somewhat beyond the scope of this manuscript, further efforts to facilitate the reproduction of analyses that require specific (older) versions of packages will be needed.

It is worth noting that \knitr provides functionality for condensing an \RM file into a conventional \R script, and vice versa. More generally, those who are comfortable working with \R scripts will find it easy to augment those scripts into \RM files, which will retain the ability to send successive \R commands to the current console.

It would be interesting to assess the extent to which students absorb the importance of reproducibility in this couse. Adding an assessment that specifically addresses reproducibility and is presented to students with a set of concrete learning objectives is something that is under consideration and a topic of future work. However it is not trivial to add material to an already busy introductory statistics curriculum, and therefore requires careful consideration of the existing material and assessments.

On a cautionary note, we remind the reader that due to the multiple uncorrected tests we ran, the claims of statistical significance made in Section \ref{sec:cor} should not be overstated.

%

\section{Conclusion}

The aforementioned \emph{Nature}~\cite{editorial2013} concludes with a call to action: ``We urge others to take note \ldots and do whatever they can to improve research reproducibility"~(p. 398).  As statistics educators, we are the members of the scientific community that are most well-suited to, and responsible for, addressing this challenge. \RM is a new technology that integrates seamlessly into existing computational work done with \R within \RStudio. With appropriate support mechanisms, introductory statistics students are receptive to its adoption. In our experience at two very different institutions with very different student bodies, \RM made a welcomed improvement to the traditional copy-and-paste workflow. Students left the course equipped with functional skills that will help them in any future quantitative endeavor. 


%
%

\section{Acknowledgements}

This work was partially supported by Project MOSAIC, US National Science Foundation (DUE-0920350).


\bibliography{references}
\vfill

\appendix

\pagebreak

\section{R Markdown Survey}
\label{sec:survey}

This survey is
part of
an ongoing research study 
to help improve the use of technology in introductory statistics courses. Responses will be merged with other assessment
data from the class to create a de-identified research dataset accessible only to the instructor, and the original
forms will be destroyed.  Only aggregate data will be included in any reports.  

The decision to participate in this study is entirely up to you.
You may refuse to take part in the study at any time without affecting
your relationship with the investigators of this study, your grade
in the course or \blind{Smith}{XX} College.  You have the right not to answer
any single question.  You are under no obligation to complete the
survey.  Your submission of the completed survey constitutes your
consent to use of the data within these constraints.

If you have any further questions about the study, at any time feel free to contact \blind{Nicholas Horton}{XX} at \blind{nhorton@smith.edu}{XX} or by telephone at \blind{413-585-3688}{XX}.  If you like, a summary of the results of the study will be sent to you. If you have any other concerns about your rights as a research participant that have not been answered by the investigators, you may contact \blind{Phil Peake}{XX}, Co-chair of the \blind{Smith}{XX} College Institutional Review Board at \blind{(413) 585-3914}{XX}.

\vspace{.2in}

\subsubsection*{Your Name: \underline{\hspace*{2.5in}}}

\vspace{.2in}

\subsubsection*{Background}
\begin{enumerate}
	\item How often had you used R \emph{prior} to taking this course (circle one)?
	\begin{center}
		had never heard of it \qquad never \qquad infrequently \qquad a few times \qquad frequently
	\end{center}
	\item To which of the following markup languages had you been exposed \emph{prior} to taking this course (circle all that apply)?
	\begin{center}
		HTML \qquad CSS \qquad XML \qquad \LaTeX \qquad Wikipedia (editing)
	\end{center}
\end{enumerate}

\subsubsection*{R Markdown}

Please indicate the response that most closely matches your attitude towards each of the following statements.

\begin{enumerate}
	\item I find the R Markdown syntax to be simple and understandable.
	\begin{center}
		no opinion \qquad strongly disagree \qquad disagree \qquad indifferent \qquad agree \qquad strongly agree
	\end{center}
	\item When my Markdown document does not compile, I know how to go about fixing it.
	\begin{center}
		no opinion \qquad strongly disagree \qquad disagree \qquad indifferent \qquad agree \qquad strongly agree
	\end{center}
	\item I am frequently frustrated by R Markdown when doing my homework.
	\begin{center}
		no opinion \qquad strongly disagree \qquad disagree \qquad indifferent \qquad agree \qquad strongly agree
	\end{center}	
	\item I am frequently frustrated by R when doing my homework.
	\begin{center}
		no opinion \qquad strongly disagree \qquad disagree \qquad indifferent \qquad agree \qquad strongly agree
	\end{center}	
	\item R Markdown makes my homework easier to read and understand.
	\begin{center}
		no opinion \qquad strongly disagree \qquad disagree \qquad indifferent \qquad agree \qquad strongly agree
	\end{center}
	\item I would rather copy and paste my results (plots, tables, and numbers) into a word processing program (e.g. Word).
	\begin{center}
		no opinion \qquad strongly disagree \qquad disagree \qquad indifferent \qquad agree \qquad strongly agree
	\end{center}
	\item I resent being forced to use R Markdown. It should be my choice how I prepare my homework.
	\begin{center}
		no opinion \qquad strongly disagree \qquad disagree \qquad indifferent \qquad agree \qquad strongly agree
	\end{center}
	\item I found R Markdown to be frustrating at first, but now I've got the hang of it.
	\begin{center}
		no opinion \qquad strongly disagree \qquad disagree \qquad indifferent \qquad agree \qquad strongly agree
	\end{center}
	\item R Markdown makes it easier for me to complete my homework.
	\begin{center}
		no opinion \qquad strongly disagree \qquad disagree \qquad indifferent \qquad agree \qquad strongly agree
	\end{center}
	\item R Markdown makes it more difficult for me to complete my homework.
	\begin{center}
		no opinion \qquad strongly disagree \qquad disagree \qquad indifferent \qquad agree \qquad strongly agree
	\end{center}
	\item I wish I had received a more thorough introduction to the logic and features of R Markdown.
	\begin{center}
		no opinion \qquad strongly disagree \qquad disagree \qquad indifferent \qquad agree \qquad strongly agree
	\end{center}
\end{enumerate}

\pagebreak

\section{Introducing R Markdown in class}

\begin{figure}[ht]
\centering
\begin{minipage}[b]{0.45\linewidth}
\includegraphics{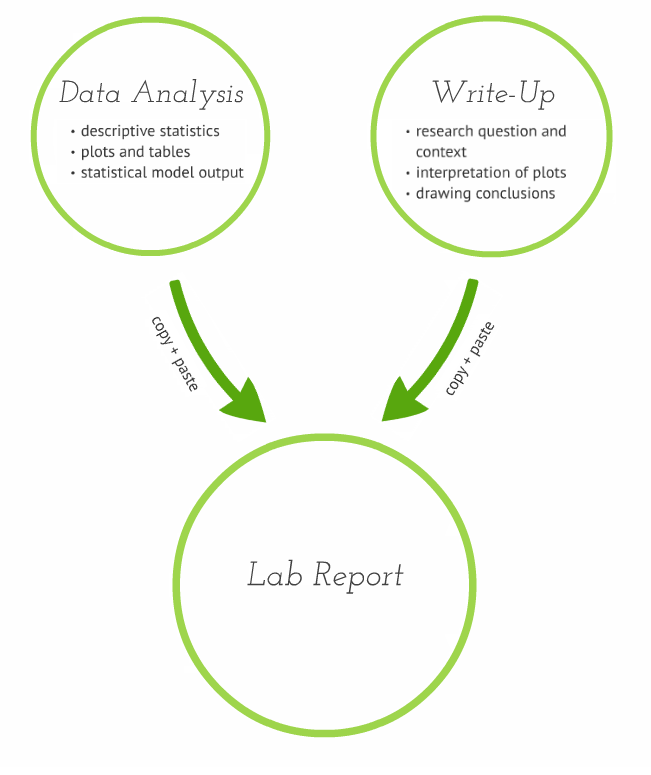}
\end{minipage}
\quad
\begin{minipage}[b]{0.45\linewidth}
\includegraphics{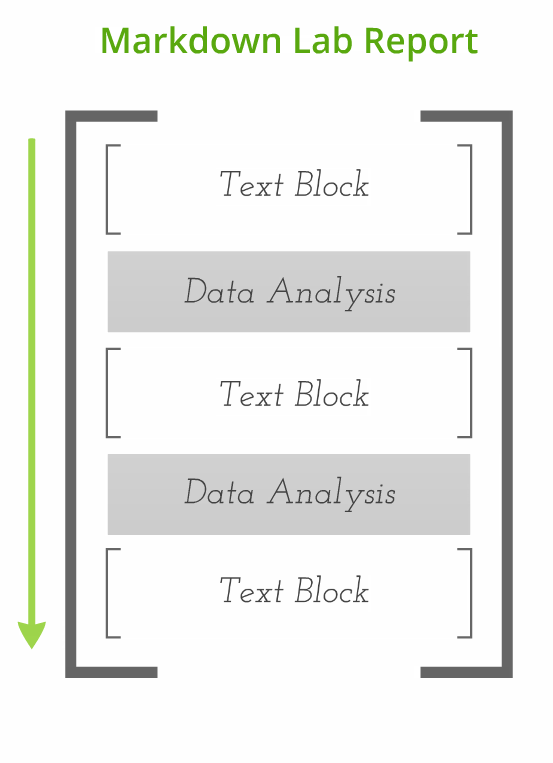}
\end{minipage}
\caption{The traditional workflow, characterized by a separation between the data analysis and the interpretation that are then fused together by copy-and-paste.  By contast, the \RM workflow integrates these two components into a single document.}
\label{fig:prezi}
\end{figure}

A Prezi introducing the features of \RM and its use in lab reports can be found at:
 \blind{\url{http://prezi.com/dvmgx17e_was/reproducible/?utm_campaign=share&utm_medium=copy}}{[url omitted for blinding purposes]}.  Figure~\ref{fig:prezi} provides two sample slides, diagramming the difference between the two workflows.

\section{Sample assignment and solution}
\label{sec:sample}

A sample lab assignment and student solution is included below.

\pagebreak



\section*{Supplementary material}

# Lab 1: Introduction to data

Some define Statistics as the field that focuses on turning information into knowledge. The first step in that process is to summarize and describe the raw information - the data. In this lab, you will gain insight into public health by generating simple graphical and numerical summaries of a data set collected by the Centers for Disease Control and Prevention (CDC). As this is a large data set, along the way you'll also learn the indispensable skills of data processing and subsetting.

## Template for lab report

Before you begin the lab, download the lab report template. This template makes it very simple to include code and output in your write up from within RStudio as well as ensuring reproducibility of your results.

```
download.file("http://stat.duke.edu/courses/Summer13/sta104.01-1/labs/lab1.Rmd",
    destfile = "lab1.Rmd")
```

Click on the file called `lab1.Rmd` under the Files tab on the bottom right pane of your RStudio window. Insert your team name, name of the "author" for the week, and the names of the "discussants" (other team members present in lab today). Use the allotted spaces to enter your responses. For questions that require R code or a plot, space has been provided for you to enter the relevant code.

## Getting started

The Behavioral Risk Factor Surveillance System (BRFSS) is an annual telephone survey of 350,000 people in the United States. As its name implies, the BRFSS is designed to identify risk factors in the adult population and report emerging health trends. For example, respondents are asked about their diet and weekly physical activity, their HIV/AIDS status, possible tobacco use, and even their level of healthcare coverage. The BRFSS Web site (*http://www.cdc.gov/brfss*) contains a complete description of the survey, including the research questions that motivate the study and many interesting results derived from the data.

We will focus on a random sample of 20,000 people from the BRFSS survey conducted in 2000. While there are over 200 variables in this data set, we will work with a small subset.

We begin by loading the data set of 20,000 observations into the R workspace. After launching RStudio, enter the following command.

```
source("http://www.openintro.org/stat/data/cdc.R")
```

The data set `cdc` that shows up in your workspace is a *data matrix*, with each row representing a *case* and each column representing a *variable*. R calls this data format a *data frame*, which is a term that will be used throughout the labs.

To view the names of the variables, type the command

```
names(cdc)
```

---

This returns the names `genhlth`, `exerany`, `hlthplan`, `smoke100`, `height`, `weight`, `wtdesire`, `age`, and `gender`. Each one of these variables corresponds to a question that was asked in the survey. For example, for `genhlth`, respondents were asked to evaluate their general health, responding either excellent, very good, good, fair or poor. The `exerany` variable indicates whether the respondent exercised in the past month (1) or did not (0). Likewise, `hlthplan` indicates whether the respondent had some form of health coverage (1) or did not (0). The `smoke100` variable indicates whether the respondent had smoked at least 100 cigarettes in her lifetime. The other variables record the respondent's `height` in inches, `weight` in pounds as well as their desired weight, `wtdesire`, `age` in years, and `gender`.

> **Exercise 1** How many cases are there in this data set? How many variables? For each variable, identify its data type (e.g. categorical, discrete).

We can have a look at the first few entries (rows) of our data with the command

```
head(cdc)
```

and similarly we can look at the last few by typing

```
tail(cdc)
```

You could also look at *all* of the data frame at once by typing its name into the console, but that might be unwise here. We know `cdc` has 20,000 rows, so viewing the entire data set would mean flooding your screen. It's better to take small peeks at the data with `head`, `tail` or the subsetting techniques that you'll learn in a moment.

## Summaries and tables

The BRFSS questionnaire is a massive trove of information. A good first step in any analysis is to distill all of that information into a few summary statistics and graphics. As a simple example, the function `summary` returns a numerical summary: minimum, first quartile, median, mean, second quartile, and maximum. For `weight` this is

```
summary(cdc$weight)
```

R also functions like a very fancy calculator. If you wanted to compute the interquartile range for the respondents' weight, you would look at the output from the summary command above and then enter

```
190 - 140
```

R also has built-in functions to compute summary statistics one by one. For instance, to calculate the mean, median, and variance of `weight`, type

```
mean(cdc$weight)
var(cdc$weight)
median(cdc$weight)
```

While it makes sense to describe a quantitative variable like `weight` in terms of these statistics, what about categorical data? We would instead consider the sample frequency or relative frequency distribution. The

function `table` does this for you by counting the number of times each kind of response was given. For example, to see the number of people who have smoked 100 cigarettes in their lifetime, type

```
table(cdc$smoke100)
```

or instead look at the relative frequency distribution by typing

```
table(cdc$smoke100)/20000
```

Notice how R automatically divides all entries in the table by 20,000 in the command above. This is similar to something we observed in the last lab; when we multiplied or divided a vector with a number, R applied that action across entries in the vectors. As we see above, this also works for tables. Next, we make a bar plot of the entries in the table by putting the table inside the barplot command.

```
barplot(table(cdc$smoke100))
```

Notice what we've done here! We've computed the table of `cdc$smoke100` and then immediately applied the graphical function, `barplot`. This is an important idea: R commands can be nested. You could also break this into two steps by typing the following:

```
smoke <- table(cdc$smoke100)
barplot(smoke)
```

Here, we've made a new object, a table, called `smoke` (the contents of which we can see by typing `smoke` into the console) and then used it in as the input for `barplot`. The special symbol <- performs an *assignment*, taking the output of one line of code and saving it into an object in your workspace. This is another important idea that we'll return to later.

> **Exercise 2** Create a numerical summary for `height` and `age`, and compute the interquartile range for each.

> **Exercise 3** Compute the relative frequency distribution for `gender` and `genhlth`. How many males are in the sample? What proportion of the sample reports being in excellent health?

The `table` command can be used to tabulate any number of variables that you provide. For example, to examine which participants have smoked across each gender, we could use the following.

```
table(cdc$gender, cdc$smoke100)
```

Here, we see column labels of 0 and 1. Recall that 1 indicates a respondent has smoked at least 100 cigarettes. The rows refer to gender. To create a mosaic plot of this table, we would enter the following command.

```
mosaicplot(table(cdc$gender, cdc$smoke100))
```

We could have accomplished this in two steps by saving the table in one line and applying `mosaicplot` in the next (see the table/barplot example above).

> **Exercise 4** What does the mosaic plot reveal about smoking habits and gender?

### Interlude: How R thinks about data

We mentioned that R stores data in data frames, which you might think of as a type of spreadsheet. Each row is a different observation (a different respondent) and each column is a different variable (the first is `genhlth`, the second `exerany` and so on). We can see the size of the data frame next to the object name in the workspace or we can type

```
dim(cdc)
```

which will return the number of rows and columns. Now, if we want to access a subset of the full data frame, we can use row-and-column notation. For example, to see the sixth variable of the 567th respondent, use the format

```
cdc[567, 6]
```

which means we want the element of our data set that is in the 567th row (meaning the 567th person or observation) and the 6th column (in this case, weight). We know that `weight` is the 6th variable because it is the 6th entry in the list of variable names

```
names(cdc)
```

To see the weights for the first 10 respondents we can type

```
cdc[1:10, 6]
```

In this expression, we have asked just for rows in the range 1 through 10. R uses the ":" to create a range of values, so 1:10 expands to 1, 2, 3, 4, 5, 6, 7, 8, 9, 10. You can see this by entering

```
1:10
```

Finally, if we want all of the data for the first 10 respondents, type

```
cdc[1:10, ]
```

By leaving out an index or a range (we didn't type anything between the comma and the square bracket), we get all the columns. When starting out in R, this is a bit counterintuitive. As a rule, we omit the column number to see all columns in a data frame. Similarly, if we leave out an index or range for the rows, we would access all the observations, not just the 567th, or rows 1 through 10. Try the following to see the weights for all 20,000 respondents fly by on your screen

```
cdc[, 6]
```

Recall that column 6 represents respondents' weight, so the command above reported all of the weights in the data set. An alternative method to access the weight data is by referring to the name. Previously,

we typed `names(cdc)` to see all the variables contained in the cdc data set. We can use any of the variable names to select items in our data set.

```
cdc$weight
```

The dollar-sign tells R to look in data frame `cdc` for the column called `weight`. Since that's a single vector, we can subset it with just a single index inside square brackets. We see the weight for the 567[th] respondent by typing

```
cdc$weight[567]
```

Similarly, for just the first 10 respondents

```
cdc$weight[1:10]
```

The command above returns the same result as the `cdc[1:10,6]` command. Both row-and-column notation and dollar-sign notation are widely used, which one you choose to use depends on your personal preference.

### A little more on subsetting

It's often useful to extract all individuals (cases) in a data set that have specific characteristics. We accomplish this through *conditioning* commands. First, consider expressions like

```
cdc$gender == "m"
```

or

```
cdc$age > 30
```

These commands produce a series of `TRUE` and `FALSE` values. There is one value for each respondent, where `TRUE` indicates that the person was male (via the first command) or older than 30 (second command).

Suppose we want to extract just the data for the men in the sample, or just for those over 30. We can use the R function `subset` to do that for us. For example, the command

```
mdata <- subset(cdc, cdc$gender == "m")
```

will create a new data set called `mdata` that contains only the men from the `cdc` data set. In addition to finding it in your workspace alongside its dimensions, you can take a peek at the first several rows as usual

```
head(mdata)
```

This new data set contains all the same variables but just under half the rows. It is also possible to tell R to keep only specific variables, which is a topic we'll discuss in a future lab. For now, the important thing

is that we can carve up the data based on values of one or more variables.

As an aside, you can use several of these conditions together with `&` and `|`. The `&` is read "and" so that

```
m_and_over30 <- subset(cdc, cdc$gender == "m" & cdc$age > 30)
```

will give you the data for men over the age of 30. The `|` character is read "or" so that

```
m_or_over30 <- subset(cdc, cdc$gender == "m" | cdc$age > 30)
```

will take people who are men or over the age of 30 (why that's an interesting group is hard to say, but right now the mechanics of this are the important thing). In principle, you may use as many "and" and "or" clauses as you like when forming a subset.

> **Exercise 5** Create a new object called `under23_and_smoke` that contains all observations of respondents under the age of 23 that have smoked 100 cigarettes in their lifetime. Write the command you used to create the new object as the answer to this exercise.

### Quantitative data

With our subsetting tools in hand, we'll now return to the task of the day: making basic summaries of the BRFSS questionnaire. We've already looked at categorical data such as `smoke` and `gender` so now let's turn our attention to quantitative data. Two common ways to visualize quantitative data are with box plots and histograms. We can construct a box plot for a single variable with the following command.

```
boxplot(cdc$height)
```

You can compare the locations of the components of the box by examining the summary statistics.

```
summary(cdc$height)
```

Confirm that the median and upper and lower quartiles reported in the numerical summary match those in the graph. The purpose of a boxplot is to provide a thumbnail sketch of a variable for the purpose of comparing across several categories. So we can, for example, compare the heights of men and women with

```
boxplot(cdc$height ~ cdc$gender)
```

The notation here is new. The `~` character can be read "versus" or "as a function of". So we're asking R to give us a box plots of heights where the groups are defined by gender.

Next let's consider a new variable that doesn't show up directly in this data set: Body Mass Index (BMI). BMI is a weight to height ratio and can be calculated as.

$$BMI = \frac{weight\ (lb)}{height\ (in)^2} * 703^{\dagger}$$

---
[†]703 is the approximate conversion factor to change units from metric (meters and kilograms) to imperial (inches and pounds)

The following two lines first make a new object called `bmi` and then creates box plots of these values, defining groups by the variable `cdc$genhlth`.

```
bmi <- (cdc$weight/cdc$height^2) * 703
boxplot(bmi ~ cdc$genhlth)
```

Notice that the first line above is just some arithmetic, but it's applied to all 20,000 numbers in the `cdc` data set. That is, for each of the 20,000 participants, we take their weight, divide by their height-squared and then multiply by 703. The result is 20,000 BMI values, one for each respondent. This is one reason why we like R: it lets us perform computations like this using very simple expressions.

> **Exercise 6** What does this box plot show? Pick another categorical variable from the data set and see how it relates to BMI. List the variable you chose, why you might think it would have a relationship to BMI, and indicate what the figure seems to suggest.

Finally, let's make some histograms. We can look at the histogram for the age of our respondents with the command

```
hist(cdc$age)
```

Histograms are generally a very good way to see the shape of a single distribution, but that shape can change depending on how the data is split between the different bins. You can control the number of bins by adding an argument to the command. In the next two lines, we first make a default histogram of `bmi` and then one with 50 breaks.

```
hist(bmi)
hist(bmi, breaks = 50)
```

Note that you can flip between plots that you've created by clicking the forward and backward arrows in the lower right region of RStudio, just above the plots. How do these two histograms compare?

> **Exercise 7** In the last lab, when exploring how percentage of boys born varies in time (two numerical variables) we use a scatterplot. Using the same tools, the `plot` function, make a scatterplot of weight versus desired weight. Describe the relationship between these variables.

At this point, we've done a good first pass at analyzing the information in the BRFSS questionnaire. We've found an interesting association between smoking and gender, and we can say something about the relationship between people's assessment of their general health and their own BMI. We've also picked up essential computing tools – summary statistics, subsetting, and plots – that will serve us well throughout this course.

## Class survey

In the rest of this lab you will use the data from the Sta 101 classes to investigate relationships between certain types of variables of interest. You can nd a list of the variables and corresponding survey questions here.

```
download.file("http://stat.duke.edu/courses/Spring13/sta101.001/data/surveyS13.csv",
    destfile = "survey.csv")
survey = read.csv("survey.csv")
```

> **Exercise 8** Pick a numerical variable, make an appropriate plot to visualize its distribution. Briefly describe the distribution of the variable using appropriate statistics. *Hint: Use R to calculate summary statistics you might want to mention in your description.*

> **Exercise 9** Pick a categorical variable, make an appropriate plot to visualize its distribution. Briefly describe the distribution of the variable using appropriate statistics.

> **Exercise 10** Pick one numerical and one categorical variable, make an appropriate plot to visualize the relationship between these variables, and briefly describe the apparent relationship.

> **Exercise 11** Pick two categorical variables, make an appropriate plot to visualize the relationship between these variables, and briefly describe the apparent relationship.

> **Exercise 12** Pick two numerical variables, make an appropriate plot to visualize the relationship between these variables, and briefly describe the apparent relationship.

> **Exercise 13** What concepts from the textbook are covered in this lab? What concepts, if any, are not covered in the textbook? Have you seen these concepts elsewhere, e.g. lecture, textbook, previous labs, etc.? Be specific in your answer.

## List of R functions

For your convenience, a list of R functions you will commonly use in this class have been posted at on the course website under the resources tab (also linked here). If you aren't sure how to do something in R, the first thing to do is to always search the web. But some of the resources you come across might be overwhelming if they're designed for more experienced users. Please don't hesitate to ask your teammates, TAs, and the professor for help.

# Lab 1

**Name:**

## Exercises

**Load CDC data:**

```
source("http://www.openintro.org/stat/data/cdc.R")
```

**Exercise 1:**

There are 20,000 cases and nine variables. Genhlth: ordinal/categorical, Exerany: categorical, Hlthplan: categorical, Smoke100: categorical, Height: discrete numerical, Weight: discrete numerical, Wtdesire: discrete numerical, Age: discrete numerical, Gender: categorical.

**Exercise 2:**

IQR of height: 70-64=6 IQR of age: 57-31=26

```
summary(cdc$height)
```

```
##    Min. 1st Qu.  Median    Mean 3rd Qu.    Max.
##    48.0    64.0    67.0    67.2    70.0    93.0
```

```
summary(cdc$age)
```

```
##    Min. 1st Qu.  Median    Mean 3rd Qu.    Max.
##    18.0    31.0    43.0    45.1    57.0    99.0
```

**Exercise 3:**

47.8% of survey participants are male, so 20,000 x .478 = 9560 participants. 23.28% of participants report being in excellent health, which makes 20,000 x .2328 = 4657 participants.

```
table(cdc$gender)/20000
```

```
##
##       m       f
## 0.4784 0.5215
```

```
table(cdc$genhlth)/20000
```

```
##
## excellent very good      good      fair      poor
##   0.23285   0.34860   0.28375   0.10095   0.03385
```

**Exercise 4:**

The mosaic plot reveals that slightly more females than males completed the survey–which we already knew. More importantly, it reveals that a larger percentage of surveyed males have smoked more than 100 cigarettes in their lives than have surveyed females. Thus, males likely have slightly worse smoking habits than females.

**Exercise 5:**

Using nrow, I find that there are 620 rows in this object, so there are 620 respondents that meet this criteria.

```
under23_and_smoke <- subset(cdc, cdc$age < "23" &
cdc$smoke100 == "1")
```

**Exercise 6:**

With these boxplots, we can see that the worse that participants declared their health to be in, the higher their median BMIs. I would think that the amount that respondents exercise also relates to the BMI–one would assume that the more someone exercises, the lower their BMI. The box plots I created indeed show that those who had not exercised in the previous month had a slightly higher median BMI than those who had.

```
bmi <- (cdc$weight/cdc$height^2) * 703
boxplot(bmi ~ cdc$exerany, main = "BMI vs. recent exercise")
```

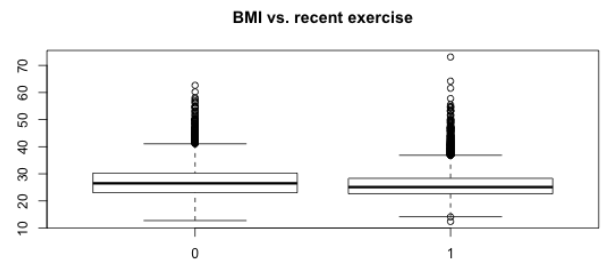

**Exercise 7:**

It appears as though the majority of respondents' desired weights are slightly below their current weights. As weight increases, desired weight generally increases as well, so the relationship is relatively strong (positive association, linear). There are a couple of outliers that seem to be inaccurate reportings (ie, desired weight of 600 and 700 lbs.), and it would be wise to disregard these data.

```
plot(cdc$wtdesire ~ cdc$weight, main = "Weight vs. desired weight")
```

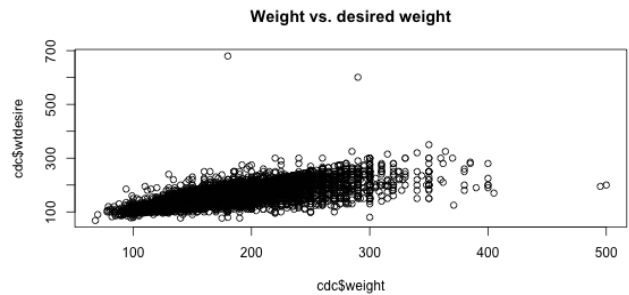

**Load survey data:**

```
download.file("http://stat.duke.edu/courses/Spring13/sta101.001/data/surveyS13.csv",
    destfile = "survey.csv")
survey = read.csv("survey.csv")
```

**Exercise 8:**

The histogram of GPAs seems to be left skewed and unimodal. Since the highest a GPA at Duke University can be is 4.0, the outliers are probably reporting mistakes and should probably not be considered with the actual data. According to the summary, the median GPA is 3.6, and using 1Q and 3Q, we can calculate that the IQR is 3.78-3.36 = .42.

```
hist(survey$gpa, breaks = 20, main = "GPAs of surveyed
students")
```

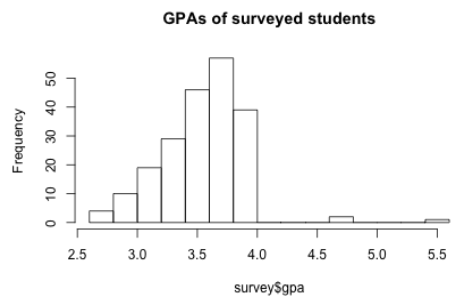

**Exercise 9:**

For this categorical variable, the most popular area of residence of surveyed students was the Southern United States (72/208 = 34.6%), followed by the Northeast US (28.3%), Western US (20.1%), Midwest US (9.6%), and finally international (7.2%).

```
table(survey$where_from)
```

```
##
##   International    US - Midwest US - Northeast       US - 
South      US - West
##              15              20              59              
72              42
```

**Exercise 10:**

With these two variables we can see that a student whose first choice was Duke goes out a median of about 2 nights per week, while the median number of nights out for students whose first choice was not Duke is one night less per week than the "yes" group.

```
boxplot(survey$go_out_times ~ survey$duke_first_choice, main
= "First choice vs. Nights out per week")
```

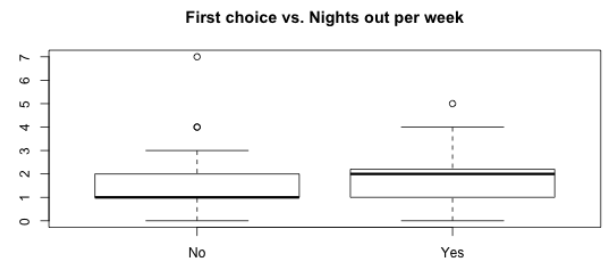

**Exercise 11:**

In general, surveyed students from regions farther away from home seem to be more homesick than students who live closer to Duke. This is evident because a larger percentage of international students, as well as students from the northeastern and western US, reported that they were homesick than did students from the midwest and southern US.

```
mosaicplot(table(survey$where_from, survey$homesick), main =
"Home location vs. Homesickness")
```

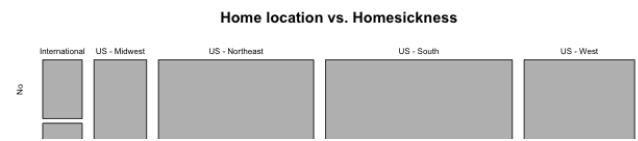

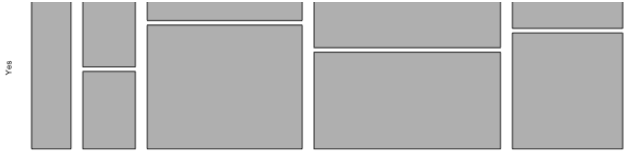

**Exercise 12:**

This scatterplot shows that there is a positive, linear association between the number of drinks it takes for a student to get drunk and the average number of drinks he/she consumes on a given night. Students who need more drinks to get drunk will generally drink more on average in a given night than students who do not need as much to get drunk, and the relationship is evident but not particularly strong. There are a few data that have higher numerical responses than the majority of the other respondants', but they follow the same general trend.

```
plot(survey$drink_amount ~ survey$drinks_to_drunk, main =
"Average drinks consumed vs. Drinks it takes to get drunk")
```

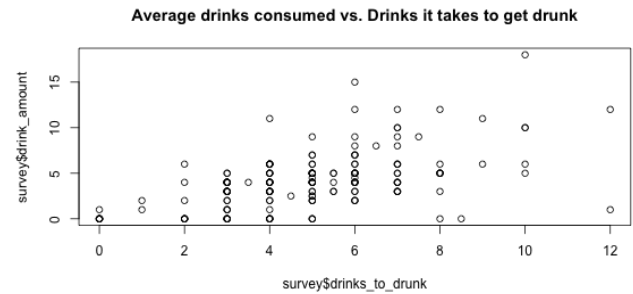

**Exercise 13:**

In this lab, we covered cases and variables, determining whether variables are numerical or categorical, how to calculate the IQR of a data set, and many different types of plots. This includes scatter plots, boxplots, mosaic plots, tables, and histograms, while using either one or two variables (both numerical and categorical) to build these plots. We also learned how to categorize these plots (for example, positive/negative association, strength, linear/nonlinear, skewness) to understand a plot's distribution or the relationship between two given variables. Additionally, we learned how to calculate information with frequency tables. We learned how to calculate a numerical summary, which is a feature of r, but was not presented in the textbook. We walked through this concept during the lecture on friday but I had not learned about it before then.



\end{document}